\documentclass[english,onecolumn,notitlepage]{revtex4-1}\usepackage{epsfig}
\usepackage{amsmath}
\usepackage{amssymb}
\usepackage{mathtools}
\usepackage[rightcaption]{sidecap}
\usepackage[figuresleft]{rotating}
\usepackage{caption}
\usepackage{float}
\usepackage{color}
\usepackage{subfigure}
\usepackage{graphicx}

\begin{document}
 \title{Stability of Majorana bound states in the presence of spin-flip scattering} 
  \author{Subhajit Pal}
 \author{Colin Benjamin} \email{colin.nano@gmail.com}\affiliation{School of Physical Sciences, National Institute of Science Education \& Research, HBNI, Jatni-752050, India }
\begin{abstract}
A popular evidence of the existence of Majorana bound states(MBS) is a quantized zero-bias conductance peak(ZBCP) which is robust to scattering by impurities, a consequence of its topological protection. In this work we examine the stability of this MBS induced ZBCP in a metal-superconductor junction in the vicinity of a spin flipper. We analytically calculate the differential charge conductance for metal-spin flipper-superconductor junctions with two distinct types of superconductors: (a) spin less $p$-wave superconductor and (b) spin-orbit-coupled s-wave superconducting wire in presence of a Zeeman field(SOCSW). We see that the quantized ZBCP remains stable in presence of spin flip scattering for metal-p-wave superconductor junction, while it loses its stability when the p-wave superconductor is replaced by a SOCSW. Further, the scattering matrix of the normal metal-p-wave superconductor junction satisfies BDI symmetry class regardless of the presence or absence of spin flip scattering. In the BDI symmetry class both Hamiltonian as well as scattering matrix satisfies particle-hole, time reversal and chiral symmetries. However, in case of normal metal-SOCSW junction the Hamiltonian as well as scattering matrix belongs to symmetry class D in absence of spin flip scattering. In the symmetry class D both Hamiltonian and the scattering matrix satisfy particle-hole symmetry relation, but do not satisfy time reversal and chiral symmetry relations. In presence of spin flip scattering the scattering matrix for SOCSW belongs to symmetry class A for which the scattering matrix does not satisfy either particle-hole or time reversal or chiral symmetry relations. The reason for ZBCP at a metal-p wave superconductor junction is perfect Andreev reflection regardless of spin flip scattering, while for a metal-SOCSW junction it is the exact cancellation between normal and Andreev reflection probabilities at zero bias and not perfect Andreev reflection, in absence of spin flip scattering. In presence of spin flip scattering, on the other hand, there is no exact cancellation at zero bias which leads to absence of quantized ZBCP for a metal-SOCSW junction. 
\end{abstract}

\maketitle
\section{{Introduction}}
In recent years Majorana bound states\cite{Ben,kit,colin} (MBS) have been studied both theoretically and experimentally in many setups. MBS are quasiparticle excitations within the superconductor at zero energy which are their own anti-particles\cite{maz}. These Majorana zero modes can be observed at the boundary with a topological superconductor\cite{kit}. Possible signatures of MBS have been reportedly seen in nanowire-superconductor hybrid device\cite{Deng,Das,Mou}, topological insulator-superconductor heterostructure\cite{xu}, atomic chains on a superconductor\cite{per,jeon} and superconducting nanowire between two ferromagnetic leads\cite{val}. Further there have been reports of the generation of Majorana states at interfaces of normal metals and spin orbit coupled superconducting wire (SOCSW)\cite{setiawan}. Our main motivation in this work is to propose a reliable check on the stability of Majorana bound states at metal-superconductor interfaces. To show this we consider a spin flipper at a metal-superconductor interface, and see that the quantized zero bias conductance peak (ZBCP) remains stable in presence of spin flip scattering for p-wave superconductor(pSc) but unfortunately it loses its stability due to spin flip scattering in case of a  SOCSW. As a consequence, ZBCP does not take universal values when a spin flipper is placed at interface of a metal-SOCSW. 

To better understand the stability of MBS to spin-flip scattering, we study the symmetry of the scattering matrix(S-Matrix) of a Metal-pSc as well as a Metal-SOCSW junction. In 1D, the Hamiltonian of a spinless pSc preserves all three symmetries- particle-hole, time reversal and chiral, implying that the Hamiltonian for a spinless pSc is in topological class BDI\cite{has}. An one dimensional normal metal-spin less pSc junction is considered in Ref.~[\onlinecite{setiawan}], and by analytically calculating the different reflection probabilities it is shown that the zero bias conductance is robustly quantized at $2e^2/h$ in the topological regime. The Hamiltonian in Ref.~[\onlinecite{setiawan}] for the normal metal-pSc junction belongs to symmetry class BDI. In Ref.~[\onlinecite{fulga}], another way to determine the symmetry class is proposed, which is using the symmetries of the scattering matrix. The symmetries of the S-Matrix can also be classified into ten symmetry classes\cite{has}. In Ref.~[\onlinecite{setiawan}] S-Matrix for normal metal-pSc junction satisfies particle hole, time reversal and chiral symmetries and thus again belongs to the symmetry class BDI which  has a topological nature in 1D. Both Hamiltonian and S-Matrix for the normal metal-pSc junction in Ref.~[\onlinecite{setiawan}] belong to the BDI symmetry class. In our work, there is a spin flipper at metal-pSc interface. We find that irrespective of spin flip scattering at the interface between metal and pSc, the zero bias S-Matrix always lies in the symmetry class BDI. The topological quantum number for class BDI is the number of negative eigenvalues of the S-Matrix, which does not change with change of parameters regardless of spin flip scattering. Thus, the topological properties of the predicted MBS at a Metal-pSc junction are robust to spin-flip scattering.

Further, in Ref.~[\onlinecite{setiawan}], similar to metal-pSc junction the metal-SOCSW junction is also studied analytically and it is seen that the zero bias conductance peak (ZBCP) is quantized. {The normal Metal-SOCSW junction Hamiltonian belongs to the symmetry class D, which in 1D is topological. The S-Matrix thus also belongs to the same symmetry class D.} In Ref.~[\onlinecite{kli}], Majorana Fermion wave functions occurring in a Metal-SOCSW junction are derived. The SOCSW Hamiltonian used in Ref.~[\onlinecite{kli}] is similar to that of Ref.~[\onlinecite{setiawan}] and the SOCSW Hamiltonian in Ref.~[\onlinecite{kli}] is also in the symmetry { class D}. {There have been two previous works which deal with a 1D linearized SOCSW Hamiltonian, Ref.~[\onlinecite{setiawan}] and Ref.~[\onlinecite{kli}]}. {In these two papers}, the linearized SOCSW Hamiltonian has two branches: interior and exterior. The interior branch Hamiltonian as well as the exterior branch Hamiltonian for Ref.~[\onlinecite{setiawan}] is in {D symmetry class}, while that in Ref.~[\onlinecite{kli}] is in symmetry class DIII. On the other hand, the S-Matrix of Normal metal-SOCSW junction of Ref.~[\onlinecite{setiawan}], {satisfies only particle-hole symmetry relation.} Therefore, both Hamiltonian and S-Matrix for normal metal-SOCSW  junction in Ref.~[\onlinecite{setiawan}] belongs to the symmetry class { D for which the topological quantum number is the determinant of the S-Matrix}. In our case, when spin flipper does not cause any spin flip, we see similar behavior as in Ref.~[\onlinecite{setiawan}], i.e., {the S-Matrix for normal metal-SOCSW junction satisfies only particle-hole symmetry relation and} there is a ZBCP which is brought out not because of perfect Andreev reflection as in normal metal-pSc junction but because of exact cancellation of normal and Andreev reflection, a fact already noticed in Ref.~[\onlinecite{setiawan}]. {For a normal metal-SOCSW junction,
we find the topological quantum number for class D (Determinant of the S-Matrix) does not change with change of parameters. But, in presence of spin flip scattering, the S-Matrix for normal metal-SOCSW junction has no symmetry and is in class A and the topological quantum number for class D, i.e., determinant of the S-Matrix changes with change of parameters}. In presence of spin flip scattering the determinant of the S-Matrix for SOCSW is complex with absolute value $1$. Thus, a 1D normal metal-SOCSW junction which purportedly hosts a MBS loses whatever topological nature it possesses when subject to spin flip scattering and as a result the ZBCP is no longer stable at a metal-SOCSW interface.

The paper is organized as follows: in the next section we first present our model for spin-flip scattering at a normal metal-pSc interface  and then at a normal metal-SOCSW interface by writing the Hamiltonian, wavefunctions and boundary conditions to determine the different scattering probabilities. Following this we discuss our results by comparing the zero-bias conductance in metal-SOCSW junction with that in metal-pSc junction in presence of spin flip scattering. We next discuss the symmetry class of the S-Matrix for both Metal-pSc junction and Metal-SOCSW junction in presence of spin flip scattering. Finally, we conclude with a brief summary of our work.

\section{{Theory}}
In this work we contrast the Majorana states arising at metal-pSc interfaces with those arising at metal-SOCSW interfaces. The exact setting we will use is shown in Figs.~1(a) and 2(a), it represents a spin flipper at $x=-a$ while at $x=0$ a $\delta$-like potential barrier which separates the normal metal from pSc or SOCSW. The regions I ($x<-a$) and II ($-a<x<0$) are metallic while for $x>0$ there is a pSc or SOCSW in the topological regime.

\subsection{{Spin flipper in the vicinity of Metal-pSc junction}} \subsubsection{ Hamiltonian} We consider a one-dimensional normal metal (NM)-normal metal (NM)-pSc junction wherein a spin flipper is embedded between two metallic regions at $x=-a$. The interface at $x=0$ (see Fig.~1) is modeled by a $\delta$-like potential barrier (strength $Z$) and the problem is solved using the BTK\cite{BTK} approach. {The normal metal is spinful and spinflip processes are confined to the normal metal.}
When a spin up/down electron with energy $E$ is incident from the metallic region, at the $x=-a$ interface, it interacts with the spin flipper through an exchange interaction which may cause a mutual spin flip. The incident electron can be reflected back to metallic region I, or transmitted to metallic region II, with spin up or down. {When this transmitted electron with spin-up is incident at $x=0$ interface it can be normally reflected back as an electron with spin-up from the interface or could be Andreev reflected, only as spin down hole back to region II. In the p-wave superconductor the spin up electron and spin down hole form one channel of transport, while the spin down electron and spin-up hole form another channel of transport for energies above the superconducting gap. There is no spin mixing in the p-wave superconductor. These are two separate channels. So, in our case the p-wave superconductor is still spinless regardless of the presence of spin flipper in its vicinity. Similar studies on spinful semiconductor coupled with spinless s-wave superconductor junction where spin flip processes are confined to the semiconductor region are dime a dozen. For e.g., in Ref.~[\onlinecite{cheng}], spin-orbit coupling is only present in the semiconductor in a superconductor-semiconductor-superconductor junction, while the superconductor is spinless.}

{In our work spin flipper is a point like magnetic impurity similar to Ref.~[\onlinecite{AJP}]. The effect of magnetic impurities on the quantum transport is vast\cite{hur,tana,samu}. This includes novel Kondo effect in which the conduction electrons scatter in a metal due to magnetic impurities. This results in a characteristic change in electrical resistivity with temperature\cite{lno}. In a diffusive normal metal-insulator-s-wave superconductor junction, normalized zero-bias conductance is suppressed by magnetic impurity scattering in diffusive normal metal for low transparent junction, while for high transparent junction zero-bias conductance is enhanced by magnetic impurity scattering\cite{yokoy}. Further, the effect of magnetic impurity on Majorana bound state is a topic of current interest\cite{gue}.}
The Hamiltonian for spin flip interaction, from Refs.~\cite{AJP,Liu,Maru,FC,ysr} is\begin{subequations}
\begin{eqnarray}
H_{Spin flipper}&=&-J\vec{s}.\vec{S'}, \\
\vec s.\vec S'&=&s_{z}S'_{z}+\frac{1}{2}(s^{-}S'^{+}+s^{+}S'^{-}),
\end{eqnarray}
\end{subequations}
$J$ denotes the strength of exchange coupling between electron's spin $\vec{s}$ and  spin $\vec{S'}$ (of spin-flipper).  $S'^{\pm} = S'_{x}\pm iS'_{y}$ are the spin raising and lowering operators for spin flipper while $s^{\pm} = s_{x}\pm is_{y}$ are the spin raising and lowering operators for electron/hole with $s_{k}=\frac{\hbar}{2}\begin{pmatrix}\sigma_{k} & 0 \\
0 & \sigma_{k} \end{pmatrix}, k=x,y,z$, with $\sigma_k$ being the usual Pauli matrices. 
{When the electron interacts with the magnetic impurity there are two possible processes, one is \textquoteleft Both no flip and mutual spin flip\textquoteright{} and another is \textquoteleft Only no flip\textquoteright. In first process, an electron (with spin $s$ and magnetic moment $m$) can scatter from the magnetic impurity (with spin $S'$ and magnetic moment $m'$) while flipping its own spin as well as flipping the spin of magnetic impurity with finite probability although there is no certainty for mutual spin flip, there can be another process without any flip. In second process, an electron (with spin $s$ and magnetic moment $m$) scatters from the magnetic impurity (with spin $S'$ and magnetic moment $m'$) without flipping either its own spin or the spin of magnetic impurity and there is no possibility of spin-flip scattering.}

The Bogoliubov-de Gennes (BdG) Hamiltonians for normal metal (NM) and $p$-wave superconductor (pSc), from Ref.~[\onlinecite{setiawan}] are-
\begin{subequations}
\begin{eqnarray}
&&H_{NM}=(-\hbar^2\partial_{x}^2/2m^{*}-\mu_{NM})\tau_{z},\\
&&H_{pSc}=(-\hbar^2\partial_{x}^2/2m^{*}-\mu_{pSc})\tau_{z}-i\Delta_{pSc}\partial_{x}\tau_{x},
\end{eqnarray}
\end{subequations}
 $\mu_{NM}$ and $\mu_{pSc}$ are the respective chemical potentials, $m^{*}$ the mass of electron, $\Delta_{pSc}\geq0$ is the p-wave pairing potential, $\tau_{\mu}=\sigma_{\mu}\otimes I$, with $I$ being $2\times 2$ identity matrix and $\sigma_{\mu} (\mu=x,y,z)$ are Pauli matrices. For simplicity, we consider $\hbar=\mu_{NM}=2m^{*}=1$. The energy spectrum are then $\varepsilon_{NM,\pm}(k)=\pm(k^2-1)$ and $\varepsilon_{pSc,\pm}(k)=\pm\sqrt{(k^2-\mu_{pSc})^2+(\Delta_{pSc}k)^2}$, respectively. {In Fig.~1 (Ref.~\cite{setiawan}) the energy spectra of the pSc for different values of $\mu_{pSc}$ is plotted. It is seen that the energy spectrum becomes gapless at $\mu_{pSc}=0$. But for nontopological regime, there is a energy gap $|\mu_{pSc}|$ in the energy spectrum.} In this work we only concentrate on the topological regime\cite{setiawan}, i.e., $\mu_{pSc}>0$. For $\mu_{pSc}>\Delta_{pSc}^2/2$, the positive energy spectrum for pSc shows the characteristic \textquotedblleft double-well\textquotedblright{} BCS structure with minima at $\epsilon_{1}=\Delta_{pSc}\sqrt{\mu_{pSc}-\Delta_{pSc}^2/4}$ for $k=\pm\sqrt{\mu_{pSc}-\Delta_{pSc}^2/2}$ and a local maximum at $\epsilon_{2}=\mu_{pSc}$ for $k=0$, as seen in Fig.~1 of Ref.~\cite{setiawan}.
\begin{figure}
\centering{\includegraphics[width=0.99\textwidth]{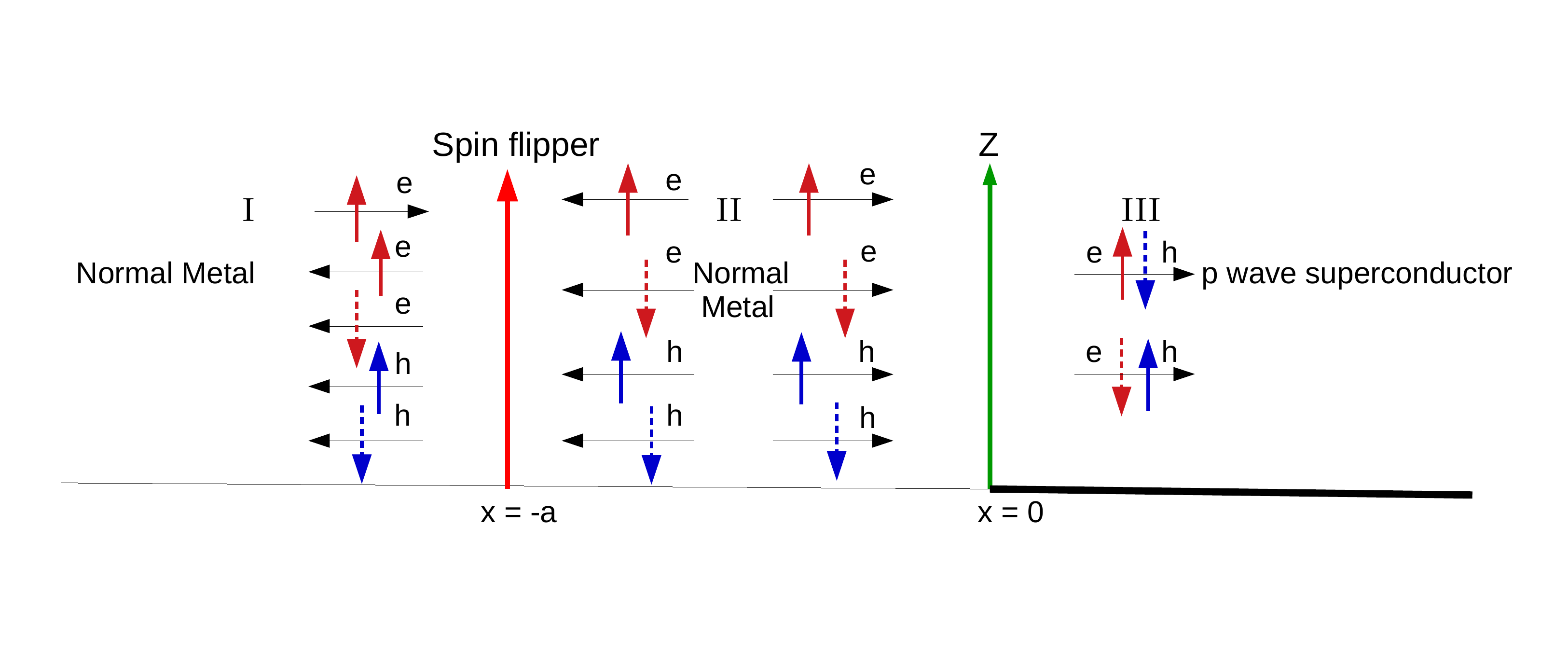}}
\caption{\small\sl  Normal metal (NM)-Normal metal (NM)-pSc junction in topological regime ($\mu_{pSc}>0$) with a spin flipper (spin $S'$, magnetic moment $m'$) at $x=-a$ and a $\delta$-like potential barrier (strength $Z$) at $x=0$. The scattering of a incident spin up electron is shown. Normal reflection, Andreev reflection and quasi-particle transmission into p-wave superconductor are represented.}
\end{figure}
\subsubsection{Wavefunctions-}
The wave functions for the different regions of our system as shown in Fig.~1 can be written  for a spin-up electron incident at $x=-a$ interface as-
\begin{widetext}
\begin{eqnarray}
\psi_{NM}^{I}(x)\!=\!\begin{pmatrix}
                    1\\
                    0\\
                    0\\
                    0
                  \end{pmatrix}e^{i(x+a)}\phi_{m'}^{S'}\!+\!r_{ee}^{\uparrow\uparrow}\begin{pmatrix}
                  1\\
                  0\\
                  0\\
                  0
                 \end{pmatrix}e^{-i(x+a)}\phi_{m'}^{S'}\!+\!r_{ee}^{\uparrow\downarrow}\begin{pmatrix}
                 0\\
                 1\\
                 0\\
                 0
                \end{pmatrix}e^{-i(x+a)}\phi_{m'+1}^{S'}\!+\!r_{eh}^{\uparrow\uparrow}{\begin{pmatrix}
                0\\
                0\\
                0\\
                1
               \end{pmatrix}}e^{i(x+a)}\phi_{m'+1}^{S'}\!+\!r_{eh}^{\uparrow\downarrow}{\begin{pmatrix}
               0\\
               0\\
               1\\
               0
              \end{pmatrix}}e^{i(x+a)}\phi_{m'}^{S'},\nonumber\\\nonumber\mbox{for $x<-a$},\\
\label{eq3}
\end{eqnarray}
\begin{eqnarray}
\psi_{NM}^{II}(x)=t_{ee}^{'\uparrow\uparrow}\begin{pmatrix}
                                     1\\
                                     0\\
                                     0\\
                                     0
                                     \end{pmatrix}e^{i(x+a)}\phi_{m'}^{S'}\!+\!t_{ee}^{'\uparrow\downarrow}\begin{pmatrix}
                                     0\\
                                     1\\
                                     0\\
                                     0
                                    \end{pmatrix}e^{i(x+a)}\phi_{m'+1}^{S'}\!+\!b_{ee}^{\uparrow\uparrow}\begin{pmatrix}
                                    1\\
                                    0\\
                                    0\\
                                    0
                                   \end{pmatrix}e^{-ix}\phi_{m'}^{S'}\!+\!b_{ee}^{\uparrow\downarrow}\begin{pmatrix}
                                   0\\
                                   1\\
                                   0\\
                                   0
                                  \end{pmatrix}e^{-ix}\phi_{m'+1}^{S'}\nonumber\\\!+\!c_{eh}^{\uparrow\uparrow}{\begin{pmatrix}
                                  0\\
                                  0\\
                                  0\\
                                  1
                                 \end{pmatrix}}e^{ix}\phi_{m'+1}^{S'}\!+\!c_{eh}^{\uparrow\downarrow}{\begin{pmatrix}
                                 0\\
                                 0\\
                                 1\\
                                 0
                                \end{pmatrix}}e^{ix}\phi_{m'}^{S'}+a_{eh}^{\uparrow\uparrow}{\begin{pmatrix}
                                0\\
                                0\\
                                0\\
                                1
                               \end{pmatrix}}e^{-i(x+a)}\phi_{m'+1}^{S'}+a_{eh}^{\uparrow\downarrow}{\begin{pmatrix}
                               0\\
                               0\\
                               1\\
                               0
                              \end{pmatrix}}e^{-i(x+a)}\phi_{m'}^{S'},\mbox{for $-a<x<0$},   
\label{eq4}
\end{eqnarray} 
\begin{eqnarray}
\psi_{pSc}(x)=t_{ee}^{\uparrow\uparrow}\begin{pmatrix}
                              \eta_{-}\\
                              0\\
                              0\\
                              1
                             \end{pmatrix}e^{ik_{-}x}\phi_{m'}^{S'}\!+\!t_{ee}^{\uparrow\downarrow}\begin{pmatrix}
                             0\\
                             \eta_{-}\\
                             1\\
                             0
                             \end{pmatrix}e^{ik_{-}x}\phi_{m'+1}^{S'}\!+\!t_{eh}^{\uparrow\uparrow}{\begin{pmatrix} \eta_{+}\\
                             0\\
                             0\\
                             1
                            \end{pmatrix}}e^{ik_{+}x}\phi_{m'+1}^{S'}\!+\!t_{eh}^{\uparrow\downarrow}{\begin{pmatrix}
                              0\\
                             \eta_{+}\\
                             1\\
                             0
                             \end{pmatrix}}e^{ik_{+}x}\phi_{m'}^{S'},\mbox{for $x>0$},
\end{eqnarray}
\end{widetext}
where $\eta_{\pm}=\frac{E+k_{\pm}^2-\mu_{pSc}}{\Delta_{pSc}k_{\pm}}$ and $\phi_{m'}^{S'}$ is the eigenspinor of spin flipper: with the $S'_{z}$ operator acting as- $S'_{z}\phi_{m'}^{S'}=m'\phi_{m'}^{S'}$, $m'$ is the magnetic moment of spin flipper. {$r_{ee}^{\uparrow\uparrow}$ and $r_{ee}^{\uparrow\downarrow}$ are the normal reflection amplitudes of an incoming electron with spin up ($\uparrow$) reflected as an electron with spin up ($\uparrow$) and spin down ($\downarrow$) respectively, while $r_{eh}^{\uparrow\uparrow}$ and $r_{eh}^{\uparrow\downarrow}$ are the Andreev reflection amplitudes of an incoming electron with spin up ($\uparrow$) reflected as a hole with spin up ($\uparrow$) and spin down ($\downarrow$) respectively.} Similarly, $t_{ee}^{\uparrow\uparrow}$, $t_{ee}^{\uparrow\downarrow}$, $t_{eh}^{\uparrow\uparrow}$, $t_{eh}^{\uparrow\downarrow}$ are the transmission amplitudes into pSc. In Eqs.~(\ref{eq3},\ref{eq4}) we approximate the wave vector in normal metal by the Fermi wave vector $k_{F}=\sqrt{2m^{*}\mu_{NM}}/\hbar=1$  (since $\hbar=\mu_{NM}=2m^{*}=1$) with $E<<E_{F}$. Wave vector's $k_{\pm}$ in p-wave superconductor are solutions of-
\begin{equation}
E^2=(k^2-\mu_{pSc})^2+(\Delta_{pSc}k)^2. 
\label{eq6}
\end{equation}
Solutions of Eq.~(\ref{eq6}) for various values of chemical potential $\mu_{pSc}>0$ (in the topological regime) with energy $E$ are mentioned in Table I of Ref.~\cite{setiawan}.

\subsubsection{Boundary conditions} Boundary conditions at $x=-a$ are- $\psi_{NM}^I(x)=\psi_{NM}^{II}(x)$ (continuity of wavefunction) and, $2i\partial_{x}\tau_{z}\psi_{NM}^{I}(x)-2i\partial_{x}\tau_{z}\psi_{NM}^{II}(x)=2iJ\vec{s}.\vec{S'}\tau_{z}\psi_{NM}^I(x)$ (discontinuity in first derivative). Boundary conditions at $x=0$ are- $\psi_{NM}^{II}(x)=\psi_{pSC}(x)$ (continuity of wavefunction) and, $(-2i\partial_{x}\tau_{z}+\Delta_{pSC}\tau_{x})\psi_{pSC}(x)+2i\partial_{x}\tau_{z}\psi_{NM}^{II}(x)=-2iZ\tau_{z}\psi_{NM}^{II}(x)$ (discontinuity in first derivative).

The action of spin-flipper Hamiltonian $\vec{s}.\vec{S'}$ on the wavefunction $\psi_{NM}^I$ via the boundary condition at $x=-a$ gives for spin up electron-
\begin{equation}
\vec s.\vec S'\tau_{z}\begin{pmatrix}
              1\\
              0\\
              0\\
              0
             \end{pmatrix}\phi_{m'}^{S'}=\frac{m'}{2}\begin{pmatrix}
             1\\
             0\\
             0\\
             0
             \end{pmatrix}\phi_{m'}^{S'}+\frac{F}{2}\begin{pmatrix}
             0\\
             1\\
             0\\
             0
             \end{pmatrix}\phi_{m'+1}^{S'},
\label{eq7}
             \end{equation}
where $F=\sqrt{(S'-m')(S'+m'+1)}$ is the spin flip probability of spin flipper. Similarly, the action of exchange operator for spin down electron gives-
\begin{equation}
\vec s.\vec S'\tau_{z}\!\begin{pmatrix}\!
              0\\
              1\\
              0\\
              0\!
             \end{pmatrix}\!\phi_{m'+1}^{S'}\!=\!-\frac{m'+1}{2}\begin{pmatrix}\!
             0\\
             1\\
             0\\
             0\!
             \end{pmatrix}\!\phi_{m'+1}^{S'}\!+\!\frac{F}{2}\begin{pmatrix}\!
             1\\
             0\\
             0\\
             0\!
             \end{pmatrix}\!\phi_{m'}^{S'}.
\label{eq8}
             \end{equation}
Further, the action of exchange operator on spin up holes gives-
\begin{equation}
\vec s.\vec S'\tau_{z}\begin{pmatrix}
              0\\
              0\\
              1\\
              0
             \end{pmatrix}\phi_{m'+1}^{S'}\!=\!\frac{m'+1}{2}\begin{pmatrix}
             0\\
             0\\
             1\\
             0
             \end{pmatrix}\phi_{m'+1}^{S'}\!-\!\frac{F}{2}\begin{pmatrix}\!
             0\\
             0\\
             0\\
             1\!
             \end{pmatrix}\phi_{m'}^{S'},
\label{eq9}
             \end{equation}
and finally the action of exchange operator on spin down holes gives-
\begin{equation}
\vec s.\vec S'\tau_{z}\begin{pmatrix}
              0\\
              0\\
              0\\
              1
             \end{pmatrix}\phi_{m'}^{S'}\!=\!-\frac{m'}{2}\begin{pmatrix}
             0\\
             0\\
             0\\
             1
             \end{pmatrix}\phi_{m'}^{S'}\!-\!\frac{F}{2}\begin{pmatrix}
             0\\
             0\\
             1\\
             0
             \end{pmatrix}\phi_{m'+1}^{S'}.
\label{eq10}
             \end{equation}
Using the above equations and solving the boundary conditions we obtain 16 equations. From these 16 equations we can compute the different scattering probabilities: $R_{ee}^{\uparrow\uparrow}=|r_{ee}^{\uparrow\uparrow}|^2$, $R_{ee}^{\uparrow\downarrow}=|r_{ee}^{\uparrow\downarrow}|^2$, $R_{eh}^{\uparrow\uparrow}=|r_{eh}^{\uparrow\uparrow}|^2$, $R_{eh}^{\uparrow\downarrow}=|r_{eh}^{\uparrow\downarrow}|^2$. Similarly, if we consider a spin down electron incident from metallic region I, we get the different scattering probabilities as follows- $R_{ee}^{\downarrow\uparrow}$, $R_{ee}^{\downarrow\downarrow}$, $R_{eh}^{\downarrow\uparrow}$, $R_{eh}^{\downarrow\downarrow}$. 

\subsection{{Spin flipper in the vicinity of metal-SOCSW junction}}
\begin{figure}[ht]
\centering{\includegraphics[width=0.99\textwidth]{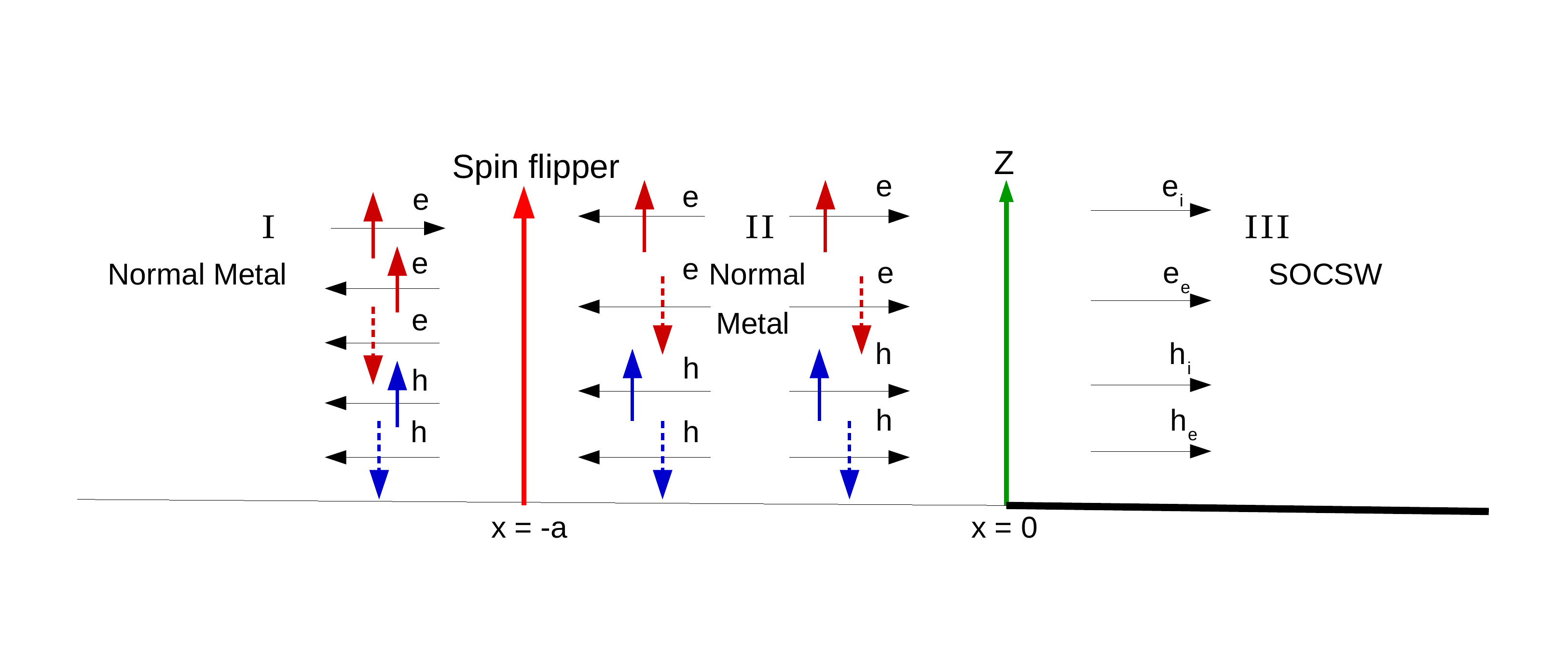}}
\caption{\small\sl Normal metal (NM)-Normal metal (NM)-SOCSW junction with a spin flipper (spin $S'$, magnetic moment $m'$) at $x=-a$ and a $\delta$-like potential barrier (strength $Z$) at $x=0$. The scattering of a incident spin-up electron is shown. Normal reflection, Andreev reflection and quasi particle transmission into SOCSW are shown.}
\end{figure}
Next, we consider a spinful normal metal (NM)-spin flipper-normal metal (NM)-Insulator (I)-SOCSW junction as shown in Fig.~2. As done before, we model the metal superconductor interface as a $\delta$-like potential barrier with strength $Z$.

\subsubsection{Hamiltonian}
Using the same convention as before, i.e., $\hbar=2m^{*}=\mu_{NM}=1$, the BdG Hamiltonians for normal metal (NM) and spin orbit coupled superconducting wire (SOCSW) can be written as\cite{setiawan}-
\begin{subequations}
\begin{eqnarray}
H_{NM}&=&(-\partial_{x}^2-1)\tau_{z},\\
H_{SOCSW}&=&-\partial_{x}^2\tau_{z}-i\beta\partial_{x}\tau_{z}\sigma_{z}\!+\!B_{Z}\sigma_{x}\!+\!\Delta_{0}\tau_{x},
\label{st1}
\end{eqnarray}
\end{subequations}
 with $\beta$ being strength of spin orbit interaction, $B_{Z}$ is Zeeman field, and $\Delta_{0}\geq0$- the proximity induced s-wave pairing potential. We also consider uniform electron masses throughout the system and fix chemical potential of the SOCSW and Zeeman interaction in the lead to be zero\cite{setiawan}. The positive branches of the energy spectrum\cite{setiawan} of SOCSW are given as $\varepsilon_{SOCSW,\pm}(k)=(k^4+\beta^2k^2+\Delta_{0}^2+B_{Z}^2\pm2\sqrt{k^4(\beta^2k^2+B_{Z}^2)+\Delta_{0}^2 B_{Z}^2})^{1/2}$. {The energy spectrum for SOCSW is shown in Fig.~4(b,c,d) of Ref.~\cite{setiawan} for nontopological ($B_{Z}<\Delta_{0}$), topological ($B_{Z}>\Delta_{0}$) and transition regimes ($B_{Z}=\Delta_{0}$). There is a gap in the energy spectrum except for $B_{Z}=\Delta_{0}$.} We focus only on the topological regime $B_{Z}>\Delta_{0}$\cite{setiawan,tew}. In the limit of strong spin orbit interaction (SOC)  ($\beta\gg B_{Z},\Delta_{0}$), the energy spectrum of the SOCSW has two branches\cite{setiawan}-(i) interior branch, (ii) exterior branch. {In Fig.~5(b) of Ref.~\cite{setiawan}  the energy spectrum for SOCSW is plotted in the limit of strong SOC. The energy spectrum about the minima at $k=0$ form the interior branches, while the energy spectrum about the minima at $k=\pm\beta$ constitute the exterior branches.} { For small energies the SOCSW Hamiltonian, Eq.~(\ref{st1}) can be linearized about these minima ($k=0$ and $k=\pm\beta$) by introducing the ansatz for the wavefunction-
\begin{subequations}
\begin{eqnarray} 
&&\psi_{SOCSW\uparrow}(x)\approx R_{\uparrow}(x)+L_{\uparrow}(x)e^{-i\beta x},\\
&&\psi_{SOCSW\downarrow}(x)\approx R_{\downarrow}(x)e^{i\beta x}+L_{\downarrow}(x),
\end{eqnarray}
\end{subequations}
where $R_{\sigma}(x)$ and $L_{\sigma}(x)$ (with $\sigma=\uparrow,\downarrow$) denote the slowly varying right moving and left moving waves respectively. By inserting this ansatz for the superconducting wavefunction in the SOCSW Hamiltonian Eq.~(\ref{st1}) and neglecting all the terms involving $e^{\pm i\beta x}$ (fast oscillating terms), we get the effective Hamiltonian corresponding to the interior and exterior branches, which is given as\cite{setiawan}-
\begin{equation}
\mathcal{H}_{SOCSW}^{(L)}=\frac{1}{2}\int dx  \Psi_{SOCSW}^{(L)}(x)^{\dagger}H_{SOCSW}^{(L)}\Psi_{SOCSW}^{(L)}(x)  
\end{equation}
where $L=i, e$ represents the interior and exterior branches respectively. The BdG Hamiltonians for the interior and exterior branches are written as-
\begin{subequations}
\begin{eqnarray}
&&H_{SOCSW}^{(i)}=-i\beta\tau_{z}\sigma_{z}\partial_{x}+B_{Z}\sigma_{x}+\Delta_{0}\tau_{x},\label{si1}\\
&&H_{SOCSW}^{(e)}=-i\beta\tau_{z}\sigma_{z}\partial_{x}+\Delta_{0}\tau_{x}.
\label{st2}
\end{eqnarray}
\label{ee}
\end{subequations}}
In addition to Ref.~[\onlinecite{setiawan}], Ref.~[\onlinecite{kli}] too linearizes the SOCSW Hamiltonian because in the limit of strong spin orbit coupling (SOC) the magnetic field and proximity-induced superconductivity are considered as small perturbations. The BdG Hamiltonian used for the superconducting nanowire in Ref.~[\onlinecite{kli}] is given by-
\begin{equation}
H_{SOCSW}^{l}=-\partial_{x}^2\tau_{z}-i\beta\sigma_{z}\partial_{x}+B_{Z}\sigma_{x}\tau_{z}+\Delta_{0}\sigma_{y}\tau_{y} 
\label{kl1}
\end{equation}
The linearized BdG Hamiltonians from Eq.~(\ref{kl1}) for the interior and exterior branches of the superconducting nanowire in the regime of strong SOC as given in Ref.~\cite{kli} is-
\begin{subequations}
\begin{eqnarray}
&&H_{SOCSW}^{l(i)}=-i\beta\sigma_{z}\partial_{x}+B_{Z}\sigma_{x}\tau_{z}+\Delta_{0}\sigma_{y}\tau_{y},\label{ki1}\\
&&H_{SOCSW}^{l(e)}=i\beta\sigma_{z}\partial_{x}+\Delta_{0}\sigma_{y}\tau_{y},
\label{kl2}
\end{eqnarray}
\end{subequations}
{If a Hamiltonian possesses particle-hole symmetry then $\tau_{y}\sigma_{y}\mathcal{H}^{*}(-k)\tau_{y}\sigma_{y}=-\mathcal{H}(k)$ ( or, $\mathcal{H}(k)=-\mathcal{H}^{*}(-k)$, or $\mathcal{H}(k)=-\tau_{x}\mathcal{H}^{*}(-k)\tau_{x}$, or $\mathcal{H}(k)=-\tau_{z}\sigma_{y}\mathcal{H}^{*}(-k)\sigma_{y}\tau_{z}$, or $\mathcal{H}(k)=-\sigma_{y}\mathcal{H}^{*}(-k)\sigma_{y}$, or $\mathcal{H}(k)=-\tau_{y}\mathcal{H}^{*}(-k)\tau_{y}$), where $k$ is the wavevector and $\tau_{x}=\begin{pmatrix}
0 & 0 & 1 & 0\\
0 & 0 & 0 & 1\\
1 & 0 & 0 & 0\\
0 & 1 & 0 & 0                                                                                                                                                                                                                                                                                                                                                                                                                                                                                                                                                                                                                                                                                                                                                                                                                                                                                                                                                                                                                                                                                                 \end{pmatrix}$, $\tau_{y}=\begin{pmatrix}
0 & 0 & -i & 0\\
0 & 0 & 0 & -i\\
i & 0 & 0 & 0\\
0 & i & 0 & 0                                                                                                                                                                                                                                                                                                                                                                                                                                                                                                                                                                                                                                                                                                                                                                                                                                                                                                                                                                                                                                                                                                 \end{pmatrix}$ and $\tau_{z}=\begin{pmatrix}
1 & 0 & 0 & 0\\
0 & 1 & 0 & 0\\
0 & 0 & -1 & 0\\
0 & 0 & 0 & -1                                                                                                                                                                                                                                                                                                                                                                                                                                                                                                                                                                                                                                                                                                                                                                                                                                                                                                                                                                                                                                                                                                 \end{pmatrix}$ are the Pauli matrices in particle-hole space, while $\sigma_{x}=\begin{pmatrix}
0 & 1 & 0 & 0\\
1 & 0 & 0 & 0\\
0 & 0 & 0 & 1\\
0 & 0 & 1 & 0                                                                                                                                                                                                                                                                                                                                                                                                                                                                                                                                                                                                                                                                                                                                                                                                                                                                                                                                                                                                                                                                                                 \end{pmatrix}$, $\sigma_{y}=\begin{pmatrix}
0 & -i & 0 & 0\\
i & 0 & 0 & 0\\
0 & 0 & 0 & -i\\
0 & 0 & i & 0                                                                                                                                                                                                                                                                                                                                                                                                                                                                                                                                                                                                                                                                                                                                                                                                                                                                                                                                                                                                                                                                                                 \end{pmatrix}$ and $\sigma_{z}=\begin{pmatrix}
1 & 0 & 0 & 0\\
0 & -1 & 0 & 0\\
0 & 0 & 1 & 0\\
0 & 0 & 0 & -1                                                                                                                                                                                                                                                                                                                                                                                                                                                                                                                                                                                                                                                                                                                                                                                                                                                                                                                                                                                                                                                                                                 \end{pmatrix}$ are Pauli matrices in spin space.}

{Similarly, if Hamiltonian possesses time reversal symmetry then $\mathcal{H}(k)=\tau_{z}\mathcal{H}^{*}(-k)\tau_{z}$ ( or, $\mathcal{H}(k)=\tau_{y}\mathcal{H}^{*}(-k)\tau_{y}$, or $\mathcal{H}(k)=\tau_{x}\mathcal{H}^{*}(-k)\tau_{x}$, or $\mathcal{H}(k)=\sigma_{y}\mathcal{H}^{*}(-k)\sigma_{y}$, or $\mathcal{H}(k)=\mathcal{H}^{*}(-k)$) and finally if Hamiltonian possesses chiral symmetry then $\mathcal{H}(k)=-\tau_{z}\mathcal{H}(k)\tau_{z}$.}

{For 1D SOCSW Hamiltonian only particle-hole symmetry holds.} The 1D SOCSW Hamiltonian in Eq.~(\ref{st1}) can be written as- 
\begin{equation}
H_{SOCSW}(k)=\begin{pmatrix}
                                            k^2+\beta k & B_{Z} & \Delta_{0} & 0\\
                                            B_{Z} & k^2-\beta k &  0 & \Delta_{0}\\
                                            \Delta_{0} & 0 & -k^2-\beta k & B_{Z}\\
                                            0 & \Delta_{0} & B_{Z} & -k^2+\beta k
                                           \end{pmatrix}\label{mt1}
\end{equation}
{ Using Eq.~(\ref{mt1}) we get-
\begin{eqnarray}
\tau_{y}\sigma_{y}H_{SOCSW}^{*}(-k)\tau_{y}\sigma_{y}={}&&\begin{pmatrix}
                                            0 & 0 & 0 & -1\\
                                            0 & 0 & 1 & 0\\
                                            0 & 1 & 0 & 0\\
                                            -1 & 0 & 0 & 0
                                           \end{pmatrix}\begin{pmatrix}
                                            k^2-\beta k & B_{Z} & \Delta_{0} & 0\\
                                            B_{Z} & k^2+\beta k &  0 & \Delta_{0}\\
                                            \Delta_{0} & 0 & -k^2+\beta k & B_{Z}\\
                                            0 & \Delta_{0} & B_{Z} & -k^2-\beta k
                                           \end{pmatrix}\begin{pmatrix}
                                            0 & 0 & 0 & -1\\
                                            0 & 0 & 1 & 0\\
                                            0 & 1 & 0 & 0\\
                                            -1 & 0 & 0 & 0
                                           \end{pmatrix}\nonumber\\
                                           ={}&&\begin{pmatrix}
                                            -k^2-\beta k & -B_{Z} & -\Delta_{0} & 0\\
                                            -B_{Z} & -k^2+\beta k &  0 & -\Delta_{0}\\
                                            -\Delta_{0} & 0 & k^2+\beta k & -B_{Z}\\
                                            0 & -\Delta_{0} & -B_{Z} & k^2-\beta k
                                           \end{pmatrix}\label{mt2}
                                           \end{eqnarray}
From Eqs.~(\ref{mt1},\ref{mt2}), we see  $\tau_{y}\sigma_{y}H_{SOCSW}^{*}(-k)\tau_{y}\sigma_{y}=-H_{SOCSW}(k)$. Similarly, we can show that time reversal symmetry does not hold as: $H_{SOCSW}(k)\neq\tau_{z}H_{SOCSW}^{*}(-k)\tau_{z}$, or $H_{SOCSW}(k)\neq\tau_{y}H_{SOCSW}^{*}(-k)\tau_{y}$, or $H_{SOCSW}(k)\neq\tau_{x}H_{SOCSW}^{*}(-k)\tau_{x}$, or $H_{SOCSW}(k)\neq\sigma_{y}H_{SOCSW}^{*}(-k)\sigma_{y}$, or $H_{SOCSW}(k)\neq H_{SOCSW}^{*}(-k)$. Finally chiral symmetry also does not hold as: $H_{SOCSW}(k)\neq-\tau_{z}H_{SOCSW}(k)\tau_{z}$. Thus, the SOCSW Hamiltonian in Eq.~(\ref{st1}) only satisfies particle-hole symmetry relation and therefore belongs to symmetry class D.}
The SOCSW Hamiltonian Eq.~(\ref{kl1}) from Ref.~[\onlinecite{kli}] also {only satisfies particle-hole symmetry relation discussed above.}
Thus, the SOCSW Hamiltonians Eq.~(\ref{st1}) used in Ref.~[\onlinecite{setiawan}] and Eq.~(\ref{kl1}) from Ref.~\onlinecite{kli} are ofcourse Hermitian, as is wont, but unlike the p-wave case (section I) which is time reversal symmetric as well as satisfies particle-hole and chiral symmetries, the SOCSW Hamiltonians {satisfy only particle-hole symmetry.}
Therefore, according to the symmetry properties of the Hamiltonian, all the Hamiltonians in  Ref.~[\onlinecite{setiawan}] and Ref.~[\onlinecite{kli}] belong to the same symmetry {class D\cite{has}}. {The exterior branches of Hamiltonian Eq.~(\ref{st2}) in Ref.~[\onlinecite{setiawan}] also satisfies particle-hole symmetry relation $\tau_{y}\sigma_{y}H_{SOCSW}^{(e)*}(-k)\tau_{y}\sigma_{y}=-H_{SOCSW}^{(e)}(k)$, and therefore belongs to D class\cite{has}.} But the exterior branch Hamiltonian Eq.~(\ref{kl2}) of Ref.~[\onlinecite{kli}] satisfies the time reversal symmetry relation $H_{SOCSW}^{l(e)}(k)=\tau_{y}H_{SOCSW}^{l(e)*}(-k)\tau_{y}$ as also the particle-hole symmetry relation $H_{SOCSW}^{l(e)}(k)=-\tau_{x}H_{SOCSW}^{l(e)*}(-k)\tau_{x}$ and thus according to the classification of symmetries of the Hamiltonian it belongs to DIII class\cite{has}. {The interior branch Hamiltonians Eqs.~(\ref{si1},\ref{ki1}) in both Ref.~[\onlinecite{setiawan}] and Ref.~[\onlinecite{kli}] satisfy only particle-hole symmetry relation, therefore belong to the same symmetry class D.} Thus, as a whole the linearized Hamiltonians for all cases belong to symmetry {class D.}

\subsubsection{Wavefunctions} The wavefunction in the normal metal regions I and II for spin up electron incident with energy $E$ is as mentioned in Eqs.~(\ref{eq3},\ref{eq4}). { For SOCSW the spinors for the interior and exterior branches are described in terms of the slowly varying left and right moving waves as $\Psi_{SOCSW}^{(i)}=\begin{pmatrix}
R_{\uparrow}(x)\\
L_{\downarrow}(x)\\
L_{\downarrow}^{\dagger}(x)\\
-R_{\uparrow}^{\dagger}(x)                                                                                                                                                    \end{pmatrix}$ and $\Psi_{SOCSW}^{(e)}=\begin{pmatrix}
L_{\uparrow}(x)\\
R_{\downarrow}(x)\\
R_{\downarrow}^{\dagger}(x)\\
-L_{\uparrow}^{\dagger}(x)                                                                                                                                                    \end{pmatrix}$. After diagonalizing the Hamiltonians in {Eqs.~(\ref{si1},\ref{st2})}, we will get the eigenfunctions for both interior and exterior branches.}
The wavefunction in the SOCSW is sum of solutions for exterior as well as an interior branch (Eq.~(\ref{ee})) and in the topological regime can be written as\cite{setiawan}-
\begin{widetext}
\begin{equation}
\psi_{SOCSW}(x)\!=\!t_{1}^{(i)}\begin{pmatrix}
                                             -u_{-}\phi_{m'}^{S'}\\
                                             -v_{-}\phi_{m'+1}^{S'}\\
                                             v_{-}\phi_{m'+1}^{S'}\\
                                             u_{-}\phi_{m'}^{S'}
                                            \end{pmatrix}e^{ik_{-}^{(i)}x}\!+\!t_{2}^{(i)}\begin{pmatrix}\!
                                                                            u_{+}\phi_{m'}^{S'}\\
                                                                            v_{+}\phi_{m'+1}^{S'}\\
                                                                            v_{+}\phi_{m'+1}^{S'}\\
                                                                            u_{+}\phi_{m'}^{S'}\!
                                                                      \end{pmatrix}e^{ik_{+}^{(i)}x}\!+\!t_{1}^{(e)}\begin{pmatrix}\!
                                                                       v_{0}\phi_{m'}^{S'}\\
                                                                       0\\
                                                                       u_{0}\phi_{m'+1}^{S'}\\
                                                                       0\!
                                                                      \end{pmatrix}e^{i(k_{0}^{(e)}\!-\!\beta)x}\!+\!t_{2}^{(e)}\begin{pmatrix}\!
                                                                      0\\
                                                                      u_{0}\phi_{m'+1}^{S'}\\
                                                                      0\\
                                                                      v_{0}\phi_{m'}^{S'}\!
                                                                      \end{pmatrix}e^{i(k_{0}^{(e)}+\beta)x}\label{eqq13}
                                                                      \end{equation}
                                                                      \end{widetext}
where the first two terms on the right-hand side denote contributions from the interior branches, while the other two terms give contributions from the exterior branches. $t_{(1,2)}^{(i)}$ and $t_{(1,2)}^{(e)}$ are the transmission amplitudes into SOCSW. 
In Eq.~(\ref{eqq13}), for $E\geq|\Delta_{\lambda}|$, $u_{\lambda}^2=\frac{E+(E^2-\Delta_{\lambda}^2)^{1/2}}{2E}$ and $u_{\lambda}^2+v_{\lambda}^2=1$, while for $0\leq E<|\Delta_{\lambda}|$, $u_{\lambda}^2=\frac{E+i(\Delta_{\lambda}^2-E^2)^{1/2}}{2|\Delta_{\lambda}|}$ and $u_{\lambda}^2+v_{\lambda}^2=\frac{E}{|\Delta_{\lambda}|}$, where $\lambda=\pm,0$, and $\Delta_{\pm}=\Delta_{0}\pm B_{Z}$. The wave vectors in Eq.~(\ref{eqq13}) are $k_{\pm}^{(i)}=\frac{(E^2-\Delta_{\pm}^2)^{1/2}}{\beta}$ for the interior branch, and $k_{0}^{(e)}=\frac{(E^2-\Delta_{0}^2)^{1/2}}{\beta}$ for the exterior branch. {In the limit of strong SOC ($\beta\gg B_{Z}$, $\Delta_{0}$), we neglect terms proportional to $k_{+}^{(i)}, k_{-}^{(i)}, k_{0}^{(e)}\ll1$ in our calculation.}

\subsubsection{Boundary conditions} The boundary conditions at $x=-a$ are- $\psi_{NM}^I(x)=\psi_{NM}^{II}(x)$ (continuity of wavefunction) and, $2i\partial_{x}\tau_{z}\psi_{NM}^{I}(x)-2i\partial_{x}\tau_{z}\psi_{NM}^{II}(x)=2iJ\vec{s}.\vec{S'}\tau_{z}\psi_{NM}^I(x)$ (discontinuity in first derivative). The boundary conditions at $x=0$ are- $\psi_{NM}^{II}(x)=\psi_{SOCSW}(x)$ (continuity of wavefunction) and, $(-2i\partial_{x}\tau_{z}+\beta\tau_{z}\sigma_{z})\psi_{SOCSW}(x)+2i\partial_{x}\tau_{z}\psi_{NM}^{II}(x)=-2iZ\tau_{z}\psi_{NM}^{II}(x)$ (discontinuity in first derivative). Substituting the wavefunctions in the above boundary conditions we get 16 equations. Solving the 16 equations we get different normal and Andreev reflection probabilities: $R_{ee}^{\uparrow\uparrow}=|r_{ee}^{\uparrow\uparrow}|^2$, $R_{ee}^{\uparrow\downarrow}=|r_{ee}^{\uparrow\downarrow}|^2$, $R_{eh}^{\uparrow\uparrow}=|r_{eh}^{\uparrow\uparrow}|^2$, $R_{eh}^{\uparrow\downarrow}=|r_{eh}^{\uparrow\downarrow}|^2$. 
Similarly, if we consider a spin down electron incident from normal metal (region I), we can easily calculate the different reflection probabilities as: $R_{ee}^{\downarrow\uparrow}=|r_{ee}^{\downarrow\uparrow}|^2$, $R_{ee}^{\downarrow\downarrow}=|r_{ee}^{\downarrow\downarrow}|^2$, $R_{eh}^{\downarrow\uparrow}=|r_{eh}^{\downarrow\uparrow}|^2$, $R_{eh}^{\downarrow\downarrow}=|r_{eh}^{\downarrow\downarrow}|^2$. We do not repeat them here but in analogy to spin up case the wavefunctions can be easily written and solved for the afore mentioned boundary conditions.

\subsection{{Differential charge conductance}}
Using the well established definitions as in Refs.~[\onlinecite{jin}, \onlinecite{kashiwaya}], we calculate the net differential charge conductance as-
\begin{eqnarray} 
&& G_{c}=G_{c}^{\uparrow}+G_{c}^{\downarrow}, \mbox{ with } G_{c}^{\uparrow}=G_{0}(1+R_{eh}^{\uparrow\uparrow}+R_{eh}^{\uparrow\downarrow}-R_{ee}^{\uparrow\uparrow}-R_{ee}^{\uparrow\downarrow})\nonumber\\
&& \mbox{and } G_{c}^{\downarrow}=G_{0}(1+R_{eh}^{\downarrow\uparrow}+R_{eh}^{\downarrow\downarrow}-R_{ee}^{\downarrow\uparrow}-R_{ee}^{\downarrow\downarrow})
\label{EQ}
\end{eqnarray}
with, $G_{0}=e^2/h$ and $G_{c}^{\uparrow}$- the differential charge conductance when spin up electron is incident from region I, while $G_{c}^{\downarrow}$- the differential charge conductance when spin down electron is incident from region I.

\section{{Results \& Analysis}} In Table I, we compare the zero-bias conductance of a NM-spin flipper-NM-I-SOCSW junction and NM-spin flipper-NM-I-pSc junction for both transparent ($Z=0$) and tunnel ($Z=3$) regimes. For spin flip case, $F\neq 0$, i.e., $S'\neq m'$, see Eqs.~(\ref{eq7}-\ref{eq10}), implying there is finite possibility for spin-flipper to flip its own spin while interacting with an electron. On the other hand, for no flip case, $F=0$, i.e., $S'=m'$. We take two different values of spin orbit coupling strength $\beta$ ($\beta=1$ and $\beta=2$) for SOCSW in second and third column of Table I. We also take two different values of $\mu_{pSc}=0.01, 0.001$, for pSc in fourth and fifth column of Table I. 
For no flip ($F=0$) case we see that the normalized zero bias conductance $G_{c}/G_{0}$ is quantized at $2$ for NM-spin flipper-NM-I-SOCSW junction, while for NM-spin flipper-NM-I-pSc junction it is quantized at $4$ regardless of other parameters like $S'$, $m'$, $J$, $Z$, $\beta$, $\mu_{pSc}$, etc. The reason for this is that the Andreev and normal reflection probabilities exactly cancel at zero bias in Eq.~(\ref{EQ}) for NM-spin flipper-NM-I-SOCSW junction while for NM-spin flipper-NM-I-pSc junction there is perfect Andreev reflection (i.e., normal reflection probabilities vanish) at zero bias and thus from Eq.~(\ref{EQ}), $G_{c}/G_{0}$ is quantized at $4$. In Table I, we have  three non-zero values of spin flip probability ($F=1,\sqrt{3},3$) for transparent junction $(Z=0)$ while two non-zero values ($F=1,3$) for tunnel junction $(Z=3)$, since the case of $F=\sqrt{3}$ is plotted in Fig.~3. 
\begin{widetext}
\begin{center}
\begin{table}[h]
\caption{Comparison of differential charge conductance at zero bias ($E=0$) in the topological regime between NM-spin flipper-NM-I-SOCSW and NM-spin flipper-NM-I-p-wave junction}  
\begin{tabular}{ |p{4.0cm}|p{4.75cm}|p{4.75cm}||p{2.1cm}|p{2.1cm}|} 
\hline
& \multicolumn{2}{|c||}{NM-spin flipper-NM-I-SOCSW} & \multicolumn{2}{|c|}{NM-spin flipper-NM-I-p-wave}\\
\hline
Parameters$\downarrow \rightarrow$ & \multicolumn{2}{|c||}{$Z=0$, $E=0$, $a=\pi$, $\Delta_{0}=0.001$, $B_{Z}=1.5\Delta_{0}$} & \multicolumn{2}{|c|}{\scriptsize $Z=0$, $E=0$, $a=\pi$, $\Delta_{pSc}=0.07$}\\  
\hline
& $G_{c}/G_{0}$ for $\beta=1$ & $G_{c}/G_{0}$ for $\beta=2$ & $G_{c}/G_{0}$ & $G_{c}/G_{0}$ \\ & & & ($\mu_{pSc}=0.001$) & ($\mu_{pSc}=0.01$)\\ 
\hline
\scriptsize No flip ($F=0, S'=m'$) & 2 & 2 & 4 & 4\\
\hline
\scriptsize Flip ($F=1, S'=-m'=1/2$) & $\frac{800+272J^2+50J^4}{400+200J^2+29J^4}$ & $\frac{2048+128J^2+50J^4}{1024+320J^2+41J^4}$ & 4 & 4\\
\hline
\scriptsize Flip ($F=\sqrt{3}, S'=-m'=3/2$) & $\frac{800+976J^2+578J^4}{400+680J^2+301J^4}$ & $\frac{2048+640J^2+578J^4}{1024+1088J^2+337J^4}$ & 4 & 4\\
\hline
\scriptsize Flip ($F=3, S'=-m'=9/2$) & $\frac{800+6928J^2+20402J^4}{400+4040J^2+10237J^4}$ & $\frac{2048+8320J^2+20402J^4}{1024+6464J^2+10345J^4}$ & 4 & 4\\
\hline\hline
Parameters$\downarrow\rightarrow$ & \multicolumn{2}{|c||}{$Z=3$, $E=0$, $a=\pi$, $\Delta_{0}=0.001$, $B_{Z}=1.5\Delta_{0}$} & \multicolumn{2}{|c|}{\scriptsize $Z=3$, $E=0$, $a=\pi$, $\Delta_{pSc}=0.07$}\\
\hline
& $G_{c}/G_{0}$ ($\beta=1$) & $G_{c}/G_{0}$ ($\beta=2$) & $G_{c}/G_{0}$ & $G_{c}/G_{0}$ \\ & & & ($\mu_{pSc}=0.001$) & ($\mu_{pSc}=0.01$)\\ 
\hline
\scriptsize No flip ($F=0, S'=m'$) & 2 & 2 & 4 & 4\\
\hline
\scriptsize Flip ($F=1, S'=-m'=1/2$) & $\frac{53792+15744J+4304J^2+480J^3+50J^4}{26896+7872J+2792J^2+336J^3+29J^4}$ & $\frac{61952+16896J+4160J^2+480J^3+50J^4}{30976+8448J+4640J^2+624J^3+41J^4}$ & 4 & 4\\
\hline
\scriptsize Flip ($F=3, S'=-m'=9/2$) &\tiny {$\frac{53792+15744J+66256J^2+9696J^3+20402J^4}{26896+7872J+38888J^2+5712J^3+10237J^4}$} & \tiny $\frac{61952+16896J+67648J^2+9696J^3+20402J^4}{30976+8448J+56864J^2+8304J^3+10345J^4}$ & 4 & 4\\
\hline
\end{tabular}  
\end{table}
\end{center}
\end{widetext}
\begin{figure*}[ht]
\centering{\includegraphics[width=0.99\textwidth]{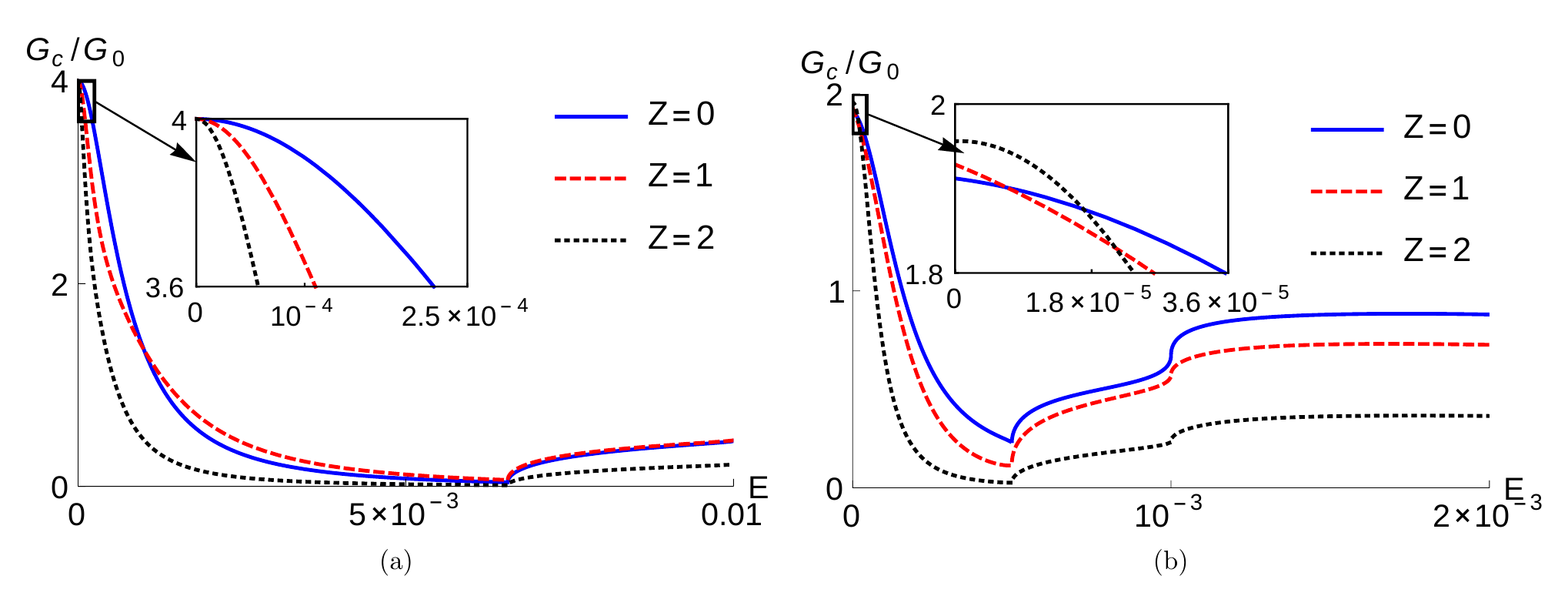}}
\caption{\small\sl Differential charge conductance in presence of spin flip scattering as a function of energy $E$ for different values of interface barrier strength $Z$ in the topological regime, (a) for NM-spin flipper-NM-I-pSc and (b) for NM-spin flipper-NM-I-SOCSW junction. Parameters are: $F=\sqrt{3}$ ($S'=-m'=3/2$), $J=1$, $\mu_{pSc}=0.01$, $\Delta_{pSc}=0.07$, $a=\pi$, $\Delta_{0}=0.001$, $\beta=0.5$, $B_{Z}=1.5\Delta_{0}$.}
\end{figure*}
We notice that for NM-spin flipper-NM-I-SOCSW junction, $G_{c}/G_{0}$ is no longer quantized and depends on the parameters $S'$, $m'$, $J$, $Z$, $\beta$, etc. However, this is not the case for NM-spin flipper-NM-I-pSc junction, wherein $G_{c}/G_{0}$ is robustly quantized and is independent of spin flip scattering. {For NM-spin flipper-NM-I-pSc junction there is perfect Andreev reflection at zero bias even in presence of spin flip scattering ($F\neq0$). Thus, from Eq.~(\ref{EQ}), zero bias conductance is quantized at $4e^2/h$ and does not depend on the parameters $S'$, $m'$, $J$, $Z$, $\mu_{pSc}$, etc.} For no flip($F=0$) case $G_{c}/G_{0}$ is robustly quantized in the topological regime for both cases, see Table I. 
Thus, in a metal-pSc junction the \textquotedblleft Majorana states\textquotedblright are not affected by presence of spin flipper, while for metal-SOCSW junction the \textquotedblleft Majorana states\textquotedblright are affected by presence of spin flipper. In Fig.~3 we plot the differential charge conductance as a function of energy $E$ for different values of interface barrier strength $Z$ in the topological regime. In Fig.~3(a) for a NM-spin flipper-NM-I-pSc junction we see that $G_{c}/G_{0}$ at $E=0$ is quantized and independent of $Z$ and spin flip probability ($F$) of spin-flipper. Thus the topological character of the zero-bias conductance peak is evident, implying the stability of Majorana state in NM-spin flipper-NM-I-pSc junction. {The differential conductance for NM-spin flipper-NM-I-pSc junction decreases with increase of $Z$ (transparency of junction). Furthermore, the width of the zero bias conductance peak (ZBCP) also decreases with increasing $Z$. A discontinuity appears in the differential conductance curve at the gap edge $\epsilon_{1}=\Delta_{pSc}\sqrt{\mu_{pSc}-\Delta_{pSc}^2/4}=0.007$, where $\Delta_{pSc}=0.07$, $\mu_{pSc}=0.01$ (topological regime) and $\mu_{pSc}>\Delta_{pSc}^2/2$ regardless of $Z$. The reason for this discontinuity is that the wave vectors $k_{\pm}$ in p-wave superconductor are different for below ($E<\epsilon_{1}$) and above ($E>\epsilon_{1}$) the gap (see Table I of Ref.~\cite{setiawan}).}
In Fig.~3(b) we do the same for NM-spin flipper-NM-I-SOCSW junction. We see that $G_{c}/G_{0}$ at $E=0$ is not quantized and depends on interface transparency $Z$. {The zero-bias differential conductance ($G_{c}/G_{0}$ at $E=0$) increases with increase of $Z$ (junction transparency). The differential conductance in NM-SOCSW case exhibits much more complex structure than that for a junction with p wave superconductor. Further, we observe a discontinuity in the differential conductance for SOCSW case at the energy corresponding to energy gaps $|\Delta_{-}|$($=0.5\Delta_{0}$) and $\Delta_{0}$. The reason for this behavior is that $u_{\lambda}$ and $v_{\lambda}$ are different for below ($0\leq E<|\Delta_{\lambda}|$) and above ($E\geq|\Delta_{\lambda}|$) the energy gap, where $\lambda=\pm,0$.}
Thus, in presence of spin flip scattering the topological character of zero-bias conductance peak seen in case of NM-spin flipper-NM-I-SOCSW junction is affected, suggesting the absence of Majorana states in such junctions.

The reason for quantized zero bias conductance for NM-spin flipper-NM-I-SOCSW junction in absence of spin-flip scattering  is exact cancellation between normal and Andreev reflection probabilities in the conductance formula (Eq.~(\ref{EQ})). Differential charge conductance is given as- $G_{c}=e^2/h(2+(R_{ee}^{\uparrow\uparrow}+R_{ee}^{\uparrow\downarrow}+R_{ee}^{\downarrow\uparrow}+R_{ee}^{\downarrow\downarrow})-(R_{eh}^{\uparrow\uparrow}+R_{eh}^{\uparrow\downarrow}+R_{eh}^{\downarrow\uparrow}+R_{eh}^{\downarrow\downarrow}))$. In no flip process and at zero energy ($E=0$), 
 $R_{ee}^{\uparrow\uparrow}+R_{ee}^{\uparrow\downarrow}+R_{ee}^{\downarrow\uparrow}+R_{ee}^{\downarrow\downarrow}=R_{eh}^{\uparrow\uparrow}+R_{eh}^{\uparrow\downarrow}+R_{eh}^{\downarrow\uparrow}+R_{eh}^{\downarrow\downarrow}=1$. Thus, normal reflection probabilities exactly cancel the Andreev reflection probabilities in conductance formula and this leads to the quantized conductance ($G_{c}=2e^2/h$) at zero bias. But, in presence of spin-flip scattering ($S'=1/2$, $m'=-1/2$, $F=1$) and for parameters $E=0$, $\beta=1$ and $B_{Z}=1.5\Delta_{0}$ from Eq.~(\ref{EQ}) we get-
\begin{eqnarray}
R_{ee}^{\uparrow\uparrow}+R_{ee}^{\uparrow\downarrow}+R_{ee}^{\downarrow\uparrow}+R_{ee}^{\downarrow\downarrow}={}&\frac{400+264J^2+33J^4+320JZ+144 J^3Z+640Z^2+352J^2Z^2+256JZ^3+256Z^4}{400+200J^2+29J^4+320JZ+112J^3Z+640Z^2+288J^2Z^2+256JZ^3+256Z^4},\label{mm1}\\\nonumber\\ 
R_{eh}^{\uparrow\uparrow}+R_{eh}^{\uparrow\downarrow}+R_{eh}^{\downarrow\uparrow}+R_{eh}^{\downarrow\downarrow}={}&\frac{400+136J^2+25J^4+320JZ+80 J^3Z+640Z^2+224J^2Z^2+256JZ^3+256Z^4}{400+200J^2+29J^4+320JZ+112J^3Z+640Z^2+288J^2Z^2+256JZ^3+256Z^4}.\label{mm2}\\\nonumber 
\end{eqnarray}
 From Eqs.~(\ref{mm1},\ref{mm2}), it is evident that- $R_{ee}^{\uparrow\uparrow}+R_{ee}^{\uparrow\downarrow}+R_{ee}^{\downarrow\uparrow}+R_{ee}^{\downarrow\downarrow}\neq R_{eh}^{\uparrow\uparrow}+R_{eh}^{\uparrow\downarrow}+R_{eh}^{\downarrow\uparrow}+R_{eh}^{\downarrow\downarrow}$. Thus, there is no exact cancellation at zero bias between normal and Andreev reflection probabilities in presence of spin flip scattering and as a consequence zero bias conductance isn't quantized at $2e^2/h$ and depends on parameters like $J$, $Z$, etc. Remarkably, by tuning parameters we can get a far greater deviation from charge conduction quantization at $2$ in case of Normal metal-SOCSW junction with spin flipper. For $S=5/2$, $m'=-5/2$, $J=3$, $Z=3$, $\beta=12$, $a=\pi$, $\Delta_{0}=0.001$, $B_{Z}=1.5\Delta_{0}$ we get $G_{c}/G_{0}=0.25$ at $E=0$, which is much smaller than the quantized value of $2$.
We have plotted the normalized conductance ($G_{c}/G_{0}$) as a function of energy $E$ for the above mentioned parameters in Fig.~4 below.
\begin{figure*}[ht]
\centering{\includegraphics[width=0.5\textwidth]{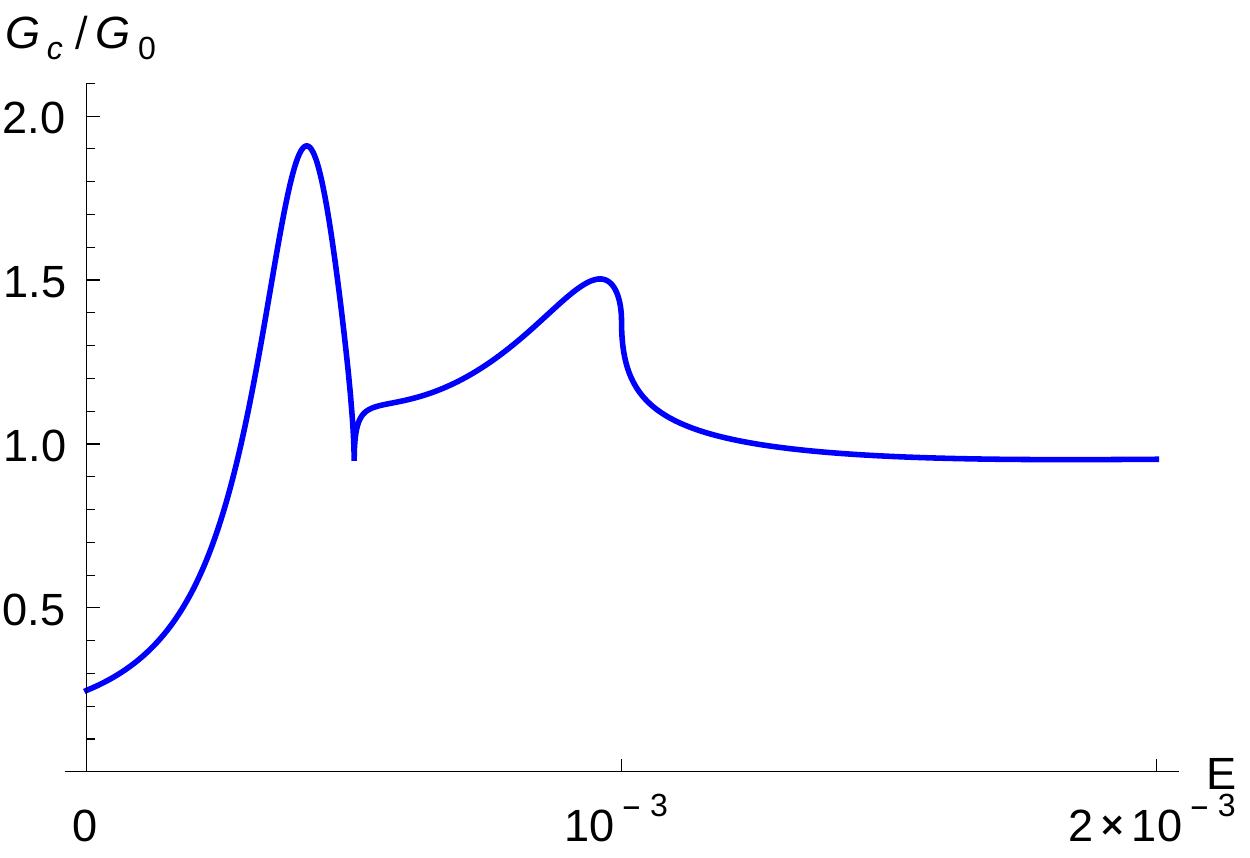}}
\caption{\small\sl Differential charge conductance in presence of spin flip scattering as a function of energy $E$ for NM-spin flipper-NM-I-SOCSW junction. Parameters are: $S=-m'=5/2$, $J=3$, $Z=3$, $a=\pi$, $\Delta_{0}=0.001$, $\beta=12$, $B_{Z}=1.5\Delta_{0}$.}
\end{figure*}
 
Next we try to find out possible reasons for this behavior by looking at the symmetries of scattering matrix and the topological quantum number. 
\subsection{Spin flip scattering and symmetry class}
To understand the deviation from topological character due to spin flip scattering in case of NM-SOCSW junction, we analyze the scattering matrix for our system, which is a $4\times4$ matrix below the gap, given by-
\begin{equation}
S_{NS}=\begin{pmatrix}
       s_{ee} & s_{eh}\\
       s_{he} & s_{hh}
       \end{pmatrix}
\end{equation}
where $s_{ee}=\begin{pmatrix}
               r_{ee}^{\uparrow\uparrow} & r_{ee}^{\uparrow\downarrow}\\
               r_{ee}^{\downarrow\uparrow} & r_{ee}^{\downarrow\downarrow}
              \end{pmatrix}$, {$s_{eh}=\begin{pmatrix}r_{eh}^{\uparrow\downarrow} & r_{eh}^{\uparrow\uparrow}\\
               r_{eh}^{\downarrow\downarrow} & r_{eh}^{\downarrow\uparrow}\end{pmatrix}$, $s_{he}=\begin{pmatrix} r_{he}^{\downarrow\uparrow} & r_{he}^{\downarrow\downarrow}\\
               r_{he}^{\uparrow\uparrow} & r_{he}^{\uparrow\downarrow}\end{pmatrix}$, $s_{hh}=\begin{pmatrix} r_{hh}^{\downarrow\downarrow} & r_{hh}^{\downarrow\uparrow}\\
               r_{hh}^{\uparrow\downarrow} & r_{hh}^{\uparrow\uparrow}\end{pmatrix}$} are all $2\times2$ block matrix. $r_{e(h)e(h)}^{\sigma\sigma'}$ is the normal reflection amplitude of an incoming electron (hole) with spin $\sigma$ reflected as a 
electron (hole) with spin $\sigma'$. Similarly, $r_{e(h)h(e)}^{\sigma\sigma'}$ is the Andreev reflection amplitude of an incoming electron (hole) with spin $\sigma$ reflected as a hole (electron) with spin $\sigma'$, $\sigma=\uparrow,\downarrow$ and $\sigma'=\uparrow,\downarrow$. Following the similar procedure as mentioned before in section II, if we consider a spin up (down) hole incident from normal metal region I, we will get different reflection amplitudes as: $r_{hh}^{\uparrow\uparrow}$ ($r_{hh}^{\downarrow\uparrow}$), $r_{hh}^{\uparrow\downarrow}$ ($r_{hh}^{\downarrow\downarrow}$), $r_{he}^{\uparrow\uparrow}$ ($r_{he}^{\downarrow\uparrow}$), $r_{he}^{\uparrow\downarrow}$ ($r_{he}^{\downarrow\downarrow}$).
\subsubsection{Normal metal-spin flipper-normal metal-p-wave superconductor junction}
For no flip process ($F=0$), and parameters $J=1$, $a=\pi$, $Z=1$, $E=0$, $\mu_{pSc}=0.01$ and $\Delta_{pSc}=0.07$, the scattering matrix for p-wave superconductor is given as-
\begin{equation}
S_{NS}=\begin{pmatrix}
        0 & 0 & 0 & -i\\
        0 & 0 & -i & 0\\
        0 & i & 0 & 0\\
        i & 0 & 0 & 0
       \end{pmatrix}\label{m2}
\end{equation}
$S_{NS}$ is an unitary matrix with $S_{NS}.S_{NS}^{\dag}=1$ and Det $S_{NS}=1$. The scattering matrix $S_{NS}$ and its determinant Det $S_{NS}$ (Det $S_{NS}=1$) do not change with change of parameters like $J$, $Z$, $a$. In Eq.~(\ref{m2}) block reflection matrices satisfy the particle-hole symmetry- $s_{ee}=s_{hh}^{*}$ and $s_{eh}=s_{he}^{*}$. Using this particle-hole symmetry we can choose a basis where all the block matrices have purely real elements. This is called Majorana basis\cite{fulga,AR} in which the scattering matrix is a real orthogonal matrix. The Majorana character of Bogoliubov quasiparticle is hidden in particle-hole basis, but becomes evident in Majorana basis. Thus, to transform the particle-hole basis into Majorana basis of the scattering matrix $S_{NS}$, we do a unitary transformation on the scattering matrix $S_{NS}$ such that the transformed scattering matrix is-
\begin{equation}
S=\Omega S_{NS}\Omega^{\dag}=\begin{pmatrix}
                                     0 & 0 & 0 & -1\\
                                     0 & 0 & -1 & 0\\
                                     0 & -1 & 0 & 0\\
                                     -1 & 0 & 0 & 0
                                    \end{pmatrix}
\end{equation}
where $\Omega=\frac{1}{\sqrt{2}}\begin{pmatrix}
                                                        1 & 0 & 1 & 0\\
                                                        0 & 1 & 0 & 1\\
                                                        -i & 0 & i & 0\\
                                                        0 & -i & 0 & i
                                                       \end{pmatrix}$, see Ref.~\cite{fulga} for details of $\Omega$.
The matrix $S$ is also an unitary matrix with Det $S=$Det $S_{NS}=1$. The transformed scattering matrix $S$ and its determinant Det $S$ remain unchanged with change of different junction parameters like $J$, $Z$, $a$. Since, the matrix $S$ satisfies $S=S^{*}$ (particle-hole symmetry), $S=S^{T}$ (time-reversal symmetry) and $S^2=1$ (chiral symmetry), thus according to the classification of symmetries of the S-Matrix for the 1D NM-$p$-wave superconductor junction is in symmetry class BDI\cite{fulga}. In 1D BDI class is topological with the topological quantum number $Q$ for class BDI defined as the number of negative eigenvalues of scattering matrix\cite{fulga,has}-
\begin{equation}
Q=\nu(r) 
\end{equation}
\begin{figure*}[ht]
\centering{\includegraphics[width=0.99\textwidth]{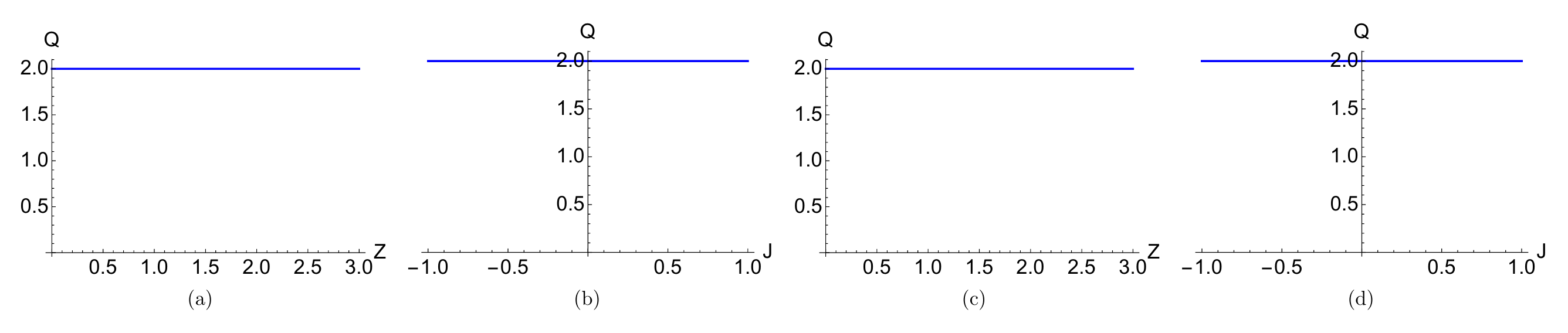}}
\caption{\small\sl (a) $Q$ in absence of spin flip scattering as a function of interface transparency $Z$, (b) $Q$  in absence of spin flip scattering as a function of exchange interaction $J$, (c) $Q$ in presence of spin flip scattering as a function of interface transparency $Z$, (d) $Q$ in presence of spin flip scattering as a function of exchange interaction $J$. Parameters are $S'=m'=1/2$ (for (a), (b)), $S'=-m'=1/2$ (for (c), (d)), $J=1$ (for (a), (c)), $Z=1$ (for (b), (d)), $\mu_{pSc}=0.01$, $\Delta_{pSc}=0.07$, $a=\pi$.}
\end{figure*}
where $\nu(r)$ is the number of negative eigenvalues of scattering matrix $S$. The eigenvalues of $S$ are- $-1$, $-1$, $1$, $1$. Thus, in no flip process topological quantum number for class BDI is $2$. In Figs.~5(a) and 5(b) we plot topological quantum number $Q$ as a function of $Z$ and $J$ respectively. Figs.~5 (a) and (b) show that $Q$ is constant at $2$ independent of $J$ and $Z$. Thus, topological quantum number $Q$ for  BDI remains robust against any change of parameters in absence of spin-flip scattering. 

Now in presence of spin-flip scattering($F\neq0$), the S-Matrix $S_{NS}$ (Eq.~\ref{m2}) for Normal Metal-p-wave superconductor junction remains identical. Similar to the preceding case of no spin-flip scattering, the S-Matrix and its determinant do not change with change of different parameters for spin-flip scattering too. Thus, in presence of spin flip scattering the 1D normal metal-$p$-wave superconductor junction also belongs to the same symmetry class BDI with topological quantum number $Q=2$. In Figs.~5(c) and 5(d) we plot topological quantum number $Q$ as a function of $Z$ and $J$ respectively in presence of spin flip scattering. We see that $Q$ remains constant at $2$ and does not change with change of $Z$ and $J$ similar to no flip process. The number of Majorana fermion at the edge of the topological superconductor is determined by the topological quantum number. In our work for the case of p-wave junction, topological quantum number for class BDI is the number of negative eigenvalues of the scattering matrix for p-wave superconductor, which is $2$ irrespective of spin flip scattering. Thus, the number of Majorana fermion is two at the edge of the superconductor.
\subsubsection{Normal metal-spin flipper-normal metal-SOCSW junction}
We do a similar analysis as was done for normal metal-p-wave superconductor junction for the normal metal-SOCSW junction. Our purpose is to understand the loss of ZBCP in presence of spin flip scattering in a NM-spin flipper-NM-I-SOCSW junction. We first analyze the case where there is no spin flip scattering. The scattering matrix, for no flip process ($S'=1/2, m'=1/2, F=0$), and for parameters $J=1$, $Z=1$, $E=0$, $\beta=0.5$, $a=\pi$, and $B_{Z}=1.5\Delta_{0}$ is-
\begin{equation}
S_{NS}=\begin{pmatrix}
0.065-0.261i & -0.543-0.043i & -0.130+0.326i & 0.696+0.152i\\
0.326-0.130i & -0.369-0.609i & 0.261-0.065i & -0.217-0.5i\\
0.217-0.5i & 0.261+0.065i & -0.369+0.609i & -0.326-0.130i\\
0.696-0.152i & 0.130+0.326i & 0.543-0.043i & 0.065+0.261i
\end{pmatrix}\label{m3} 
\end{equation}
where $S_{NS}$ is a unitary matrix with $S_{NS}.S_{NS}^{\dag}=1$ and Det $S_{NS}=-1$. The S-Matrix $S_{NS}$ however changes with change of parameters unlike the S-Matrix for Normal metal-p-wave superconductor junction. For $a=\pi/2$ it  is-
\begin{equation}
S_{NS}=\begin{pmatrix}
-0.133+0.668i & 0.594-0.096i & -0.214+0.228i & 0.255+0.107i\\
-0.214+0.228i & 0.255+0.107i &  0.659-0.173i & -0.346-0.492i\\
0.346-0.492i &  0.659+0.173i &  0.255-0.107i &  0.214+0.228i\\
0.255-0.107i & 0.214+0.228i &  -0.594-0.096i & -0.133-0.668i
\end{pmatrix}\label{m4} 
\end{equation}
where the other parameters are same as for Eq.~(\ref{m3}). As shown in Eqs.~(\ref{m3},\ref{m4}), the S-Matrix for a 1D  normal metal-SOCSW junction in absence of spin-flip scattering changes with change in junction length $a$, while its determinant (Det $S_{NS}=-1$) does not depend on $a$, in both cases $a=\pi$ and $a=\pi/2$, Det $S_{NS}=-1$. We have checked that the S-Matrix $S_{NS}$ changes with other parameters like $J$, $Z$, although its determinant remains constant at $-1$ (Det $S_{NS}=-1$). Similarly, as mentioned before, when we transform the particle-hole basis of the S-Matrix into Majorana basis by doing an unitary transformation on $S_{NS}$, the transformed S-Matrix is-
\begin{eqnarray}
\mbox{for $a=\pi$:}{}&\hspace{1cm}
S=\begin{pmatrix}
-0.109+0.087i & 0.043+0.022i & 0.848+0.391i & -0.326i\\
0.913-0.196i & -0.196-0.261i & 0.087+0.109i & 0.022-0.043i\\
-0.022-0.043i & 0.087-0.109i & -0.196+0.261i & -0.913-0.196i\\
0.326i & -0.848+0.391i & -0.043+0.022i & -0.109-0.087i
\end{pmatrix}\\\nonumber\\\nonumber\\
\mbox{ and for $a=\pi/2$ is:}{}&\hspace{1cm}
S=\begin{pmatrix}
0.127+0.148i & 0.861+0.206i & -0.028+0.087i & 0.129+0.392i\\
0.053-0.074i & -0.005-0.412i & -0.195-0.012i & -0.747+0.474i\\
0.747+0.474i & -0.195+0.012i & -0.005+0.412i & -0.053-0.074i\\
0.129-0.392i &  0.028+0.087i & -0.861+0.206i &  0.127-0.148i 
\end{pmatrix}
\end{eqnarray}
$S$ is again an unitary matrix with Det $S=$Det $S_{NS}=-1$. Similar to $S_{NS}$, the transformed S-Matrix $S$ changes with change of parameters like $J$, $Z$, $a$, but its determinant remains unchanged. {We verify that the matrix $S$ only satisfies particle-hole symmetry relation $\tau_{y}\sigma_{y}S^{*}\tau_{y}\sigma_{y}=S$, thus the S-Matrix for no spin-flip scattering in case of a 1D normal metal-SOCSW junction belongs to symmetry class D.}
\begin{figure*}[ht]
\centering{\includegraphics[width=0.99\textwidth]{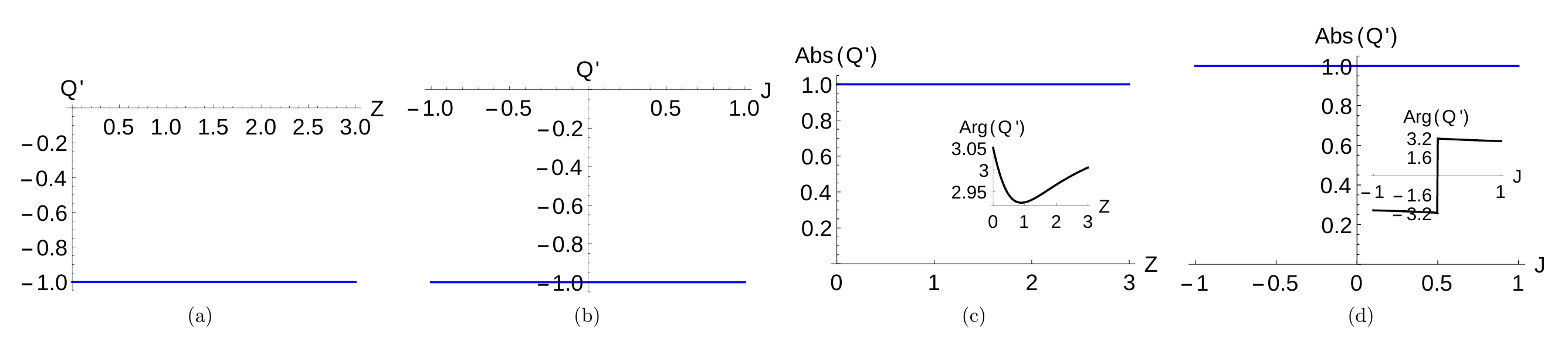}}
\caption{\small\sl (a)  $Q'$  as a function of $Z$ (with $J=1$) and (b) $Q'$ as a function of $J$ (with $Z=1$) in absence of spin-flip scattering. (c) Absolute value and Argument of complex $Q'$ as a function of $Z$ (with $J=1$) and (d) Absolute value and Argument of complex $Q'$ as a function of  $J$ (with $Z=1$) in presence of spin-flip scattering. Parameters are $S'=m'=1/2$ (for (a), (b)), $S'=-m'=1/2$ (for (c), (d)), $\Delta_{0}=0.001$, $\beta=0.5$, $a=\pi$.} 
\end{figure*}
{In 1D, symmetry class D is topological with the topological quantum number\cite{ryu} for class D defined as the determinant of the scattering matrix\cite{fulga,has}-
\begin{equation}
Q'=\mbox{Det}(S) 
\end{equation}}
{In Figs.~6(a) and (b), topological quantum number $Q'$ is plotted as a function of $Z$ and $J$ respectively. We see that the $Q'$ is constant at $-1$ and does not change with change in $Z$ and $J$. Thus, topological quantum number $Q'$ for class D remains robust against any change of parameters in absence of spin-flip scattering.}

In presence of spin-flip scattering ($S'=1/2$, $m'=-1/2$, $F=1$), and parameters $J=1$, $Z=1$, $E=0$, $\beta=0.5$ and $B_{Z}=1.5\Delta_{0}$, the S-Matrix for a 1D normal metal-SOCSW junction is given as-
\begin{eqnarray}
\mbox{$a=\frac{\pi}{2}$:}{}&\hspace{1cm}
S_{NS}=\begin{pmatrix}
-0.055+0.435i & 0.428+0.199i & -0.109+0.207i & 0.698+0.207i\\
-0.436+0.492i & -0.055+0.435i & 0.184-0.039i & -0.232-0.536i\\
-0.232-0.536i &  0.698+0.207i & 0.151-0.301i & -0.022-0.123i\\
0.184-0.039i &  -0.109+0.207i & -0.767-0.456i & 0.151-0.301i
\end{pmatrix}\\\nonumber\\\nonumber\\
\mbox{$a=\pi$:}{}&\hspace{1cm}
S_{NS}=\begin{pmatrix}
        -0.166-0.525i & -0.313+0.121i & 0.249+0.299i & 0.656+0.037i\\
        0.546+0.028i & -0.166-0.525i & 0.289-0.140i & 0.161-0.518i\\
        0.161-0.518i & 0.656+0.037i & -0.158+0.343i & -0.145-0.332i\\
        0.289-0.140i & 0.249+0.299i & 0.733-0.254i & -0.158+0.343i
       \end{pmatrix}
\end{eqnarray}
where $S_{NS}$ is a unitary matrix with $S_{NS}.S_{NS}^{\dag}=1$. The determinant of the S-Matrix for $a=\pi/2$ and $a=\pi$ are $e^{-i2.76667}$ and $e^{i2.92505}$ respectively with absolute value $1$. In contrast to no flip case, both the S-Matrix ($S_{NS}$) and its determinant (Det $S_{NS}$) change with change of parameters like $J$, $Z$. Similarly, as discussed before, we do a unitary transformation on $S_{NS}$ to transform its basis from particle-hole to Majorana. In Majorana basis the transformed scattering matrix is given as-
\begin{eqnarray}
\mbox{$a=\frac{\pi}{2}$:}{}&\hspace{1cm}
S=\begin{pmatrix}
-0.123-0.098i & 0.901+0.245i & 0.003-0.164i & -0.161+0.225i\\
-0.418-0.021i & -0.123-0.098i & -0.474+0.166i & -0.73-0.041i\\
0.739+0.041i & 0.161-0.225i & 0.219+0.231i & -0.495-0.168i\\
0.474-0.166i & -0.003+0.164i & -0.785+0.057i & 0.219+0.231i 
\end{pmatrix}\\\nonumber\\\nonumber\\
\mbox{$a=\pi$:}{}&\hspace{1cm}
S=\begin{pmatrix}
0.043-0.200i & 0.427-0.068i & 0.843-0.048i & -0.227-0.084i\\
0.929-0.253i & 0.043-0.200i & -0.141-0.093i & 0.026+0.041i\\
-0.026-0.041i & 0.227+0.084i & -0.368+0.018i & -0.885-0.143i\\
0.141+0.093i & -0.843+0.048i & 0.351+0.027i & -0.368+0.018i
\end{pmatrix}
\end{eqnarray}
where $S$ is an unitary matrix with Det $S=$Det $S_{NS}$. Similar to $S_{NS}$, its determinant Det $S_{NS}$ depends on the various junction parameters. {We have checked that the S-Matrix $S$ does not satisfy any symmetry relation and therefore belongs to class A in presence of spin flip scattering.}

{In Figs.~6(c) and 6(d) we plot both absolute value and argument of the complex $Q'$ (determinant of $S$) as a function of $Z$ and $J$ respectively for a 1D normal metal-SOCSW junction(Fig.~2). We notice that Abs($Q'$) remains constant at $1$, while the argument of the complex $Q'$ (Arg(Det $S$)) changes with $J$ and $Z$.}

\section{Conclusion}
To conclude, we have shown that zero bias quantized conductance Majorana peaks remain unaffected, in presence of spin flipper, at metal-p-wave superconductor interface while zero bias conductance peak at metal-SOCSW interface loses its quantization in presence of spin flip scattering. Further, we find that in presence of spin flip scattering a normal metal-p-wave superconductor junction belongs to the symmetry class BDI and topological quantum number for class BDI does not change with change of parameters. {But, for normal-metal-SOCSW junction the topological quantum number for class D, i.e., the determinant of the scattering matrix is complex and Det $S$ changes with change of parameters due to spin flip scattering. The S-Matrix for a  1D normal metal-SOCSW junction satisfies only particle-hole symmetry relation and belongs to topological symmetry class D in absence of spin flip scattering, while in presence of spin flip scattering the S-Matrix does not satisfy any symmetry relation and thus belongs to the non-topological class A.} The reason for ZBCP in a 1D normal metal-SOCSW junction in absence of spin-flip scattering is the exact cancellation at zero bias of the normal and Andreev reflection probabilities. However, in presence of spin flip scattering in  a 1D normal metal-SOCSW junction the exact cancellation of normal and Andreev reflection probabilities is no longer valid and as a consequence the ZBCP loses it quantization.
\section{Authors' contributions}
C.B. conceived the proposal, S.P. did the calculations on the advice of C.B., C.B. and S.P. analyzed the results and wrote the paper. Both authors reviewed the manuscript.
\section{Funding}
This work was supported by the grant ``Non-local correlations in nanoscale systems: Role of decoherence, interactions, disorder and pairing symmetry'' from SCIENCE \& ENGINEERING RESEARCH BOARD, New Delhi, Government of India, Grant No.  EMR/20l5/001836,
Principal Investigator: Dr. Colin Benjamin, National Institute of Science Education and Research, Bhubaneswar, India.
\section{Data Availability}
Data available on request from the
authors. The data that support the findings of this study are available from the corresponding author upon reasonable request.

\end{document}